\documentclass[aps,prd,12pt,notitlepage,nofootinbib,tightenlines]{revtex4-1}
\usepackage{amsmath}
\usepackage{bm}
\usepackage{times}
\usepackage{braket}
\usepackage{color}
\usepackage{epsfig}
\usepackage{slashed}
\usepackage{hyperref}
\newcommand{\beq}{\begin{eqnarray}}
\newcommand{\eeq}{\end{eqnarray}}
\newcommand{\non}{\nonumber\\ }

\newcommand{\etar}{\eta^\prime }
\newcommand{\etap}{\eta^{(\prime)} }

\newcommand{\cala}{ {\cal A} }
\newcommand{\calb}{ {\cal B} }

\newcommand{\mbs}{m_{B_s} }

\newcommand{\psl}{ P \hspace{-2.4truemm}/ }
\newcommand{\nsl}{ n \hspace{-2.2truemm}/ }
\newcommand{\vsl}{ v \hspace{-2.2truemm}/ }

\newcommand{\calh}{ {\cal H} }
\newcommand{\cals}{ {\cal S} }

\def \csb{ Chin. Sci. Bull. }

\def \epjc{ Eur. Phys. J. C }

\def \npb{  Nucl. Phys. B }
\def \plb{  Phys. Lett. B }
\def \ppnp{ Prog.Part. $\&$ Nucl. Phys. }
\def \prd{  Phys. Rev. D }
\def \prl{  Phys. Rev. Lett.  }

\def \zpc{  Z. Phys. C }
\def \jhep{ JHEP }

\definecolor{Red}{rgb}{1.,0.,0.}
\newcommand{\Red}[1]{{\color{nicered}{#1}}}
\definecolor{Blue}{rgb}{0.,0.,1.}

\definecolor{nicered}{rgb}{0.8,0.1,0.1}
\definecolor{nicegreen}{rgb}{0.1,0.5,0.1}

\hypersetup{colorlinks,citecolor=nicegreen,linkcolor=nicered}
\begin{document}

\title{\boldmath Anatomy of $B_s \to PP $ decays and effects of the next-to-leading order contributions in the perturbative QCD approach}
\author{Da-Cheng Yan$^{1}$}  \email{1019453259@qq.com}
\author{Xin Liu$^{2}$}  \email{liuxin@jsnu.edu.cn}
\author{Zhen-Jun Xiao$^{1,3}$} \email{xiaozhenjun@njnu.edu.cn}
\affiliation{$^1$ Department of Physics and Institute of Theoretical Physics,
Nanjing Normal University, Nanjing, Jiangsu 210023, P.R. China}
\affiliation{$^2$ School of Physics and Electronic Engineering, Jiangsu Normal University, Xuzhou 221116, P.R. China}
\affiliation{$^3$ Jiangsu Key Laboratory for Numerical Simulation of Large Scale Complex
Systems, Nanjing Normal University, Nanjing, Jiangsu 210023, P.R. China}
\date{\today}
\begin{abstract}
By employing the perturbative QCD (PQCD) factorization approach, we made a systematic investigation for  the CP-averaged branching
ratios and the CP-violating asymmetries of the thirteen $\bar{B}^0_s \to PP $ decays ( here $P=(\pi,K, \eta, \eta^{\prime})$)
with the inclusion of all currently known next-to-leading order (NLO) contributions, and compared our results
with the measured values or the theoretical predictions from other different approaches.
We focused on the examination of the effects of  the NLO contributions and found the following points:
(a)  for  $\bar{B}_s^0  \to (K^0 \bar{K}^0, K^+ K^-, \eta^{\prime} \eta^{\prime},\pi^- K^+)$ decays,  the NLO contributions can provide  a  $(40-150) \%$ enhancement or
a $20\%$ reduction to the corresponding  leading order (LO)  PQCD predictions for their decay rates and  result in  a much better
agreement between the PQCD predictions  and the experimental measurements;
(b) for  the pure annihilation decay $\bar{B}_s^0  \to  \pi^+\pi^-$,  the PQCD prediction still  remain consistent with the data after the inclusion
of the small NLO reduction;
(c)  the PQCD predictions for the ratio $(f_s/f_d)\cdot {\cal B}(B_s^0\to \pi^+\pi^-)/{\cal B}(B^0\to K^+\pi^-)$ and
$(f_s/f_d)\cdot {\cal B}(B_s^0\to \pi^+ K^-)/{\cal B}(B^0\to K^+\pi^-)$  become agree very well  with the measured
ones after the inclusion of   a  $40\%$  NLO reduction;
(d)   for $\bar{B}_s^0 \to K^+ K^-$ and $\bar{B}_s^0 \to \pi^- K^+$ decays, the NLO PQCD predictions
for their  CP-violating asymmetries do agree very well with the measured values in both the sign and the magnitude;
and (e) for all $\bar{B}_s^0  \to P P$ decays, we also compared our results with those
obtained in the  QCD factorization approach and  soft-collinear effective theory and discussed their  similarities and differences.
\end{abstract}
\vspace{1cm}

\maketitle

\section{Introduction}

During the past three decades, the two-body charmless hadronic $B_s \to PP$ decays ( $P$ refers to the light pseudoscalar mesons $\pi, K, \etap$)
have been studied intensively by many  authors for example in
Refs.~\cite{npb675,sun2003,chengbs09,scet06,ali07,bspipi,bs08,xiao2012,xiao14a,xiao14b,Cheng:2011qh,Chang:2014yma,Cheng:2014rfa}
and measured by CDF, Belle and LHCb Collaborations
~\cite{bellepipi,bellekk,CDFpipi,CDFpik,CDFpikcp,CDFkk,LHCbpipi,LHCbepep,LHCbpik,LHCbpikcp1,LHCbpikcp2,lhcb-18a,LHCbkkcp}.
The studies for these decays can offer us good  opportunities  to test the Standard Model (SM) and to search for the new physics (NP) beyond the SM.

Among the thirteen $B_s^0\to PP $ decays, only five of them have been observed  by  CDF \cite{CDFpipi,CDFpik,CDFkk,CDFpikcp},
Belle \cite{bellepipi,bellekk} and LHCb Collaboration \cite{LHCbpipi,LHCbepep,LHCbpik,LHCbpikcp1,LHCbpikcp2,lhcb-18a,LHCbkkcp}.
The measured values of the branching ratios
and/or the  CP-violating asymmetries are  collected in Tables \ref{tab:brexp} and  \ref{tab:acpexp}  .
Of   course,  more measurements with higher  precision for these decays in the LHCb and Belle-II experiments are expected in the
following years  ~\cite{belle-2,lhcb-2,bfac,hfag2018,pdg2018}.
On the theory side,   $B_s^0\to PP $  decays have been studied  by employing rather different kinds of theoretical approaches:
such as the generalized factorization approach \cite{chenbs99,xiaobs01,aag1},  the QCD factorization ( QCDF) approach~ \cite{prl99,npb591},
the soft-collinear effective theory (SCET) \cite{scet01,scet02}
and the perturbative QCD (PQCD) factorization approach~ \cite{pqcd1,pqcd2,li2003}.
Although there exist many clear differences between  the theoretical predictions from rather different approaches,  specifically for the pattern and magnitudes of the
CP-violating asymmetries, they are  generally consistent with each other for the branching ratios within still large theoretical errors.

In the framework of the PQCD factorization approach,  the $B_s \to PV$ and $B_s\to VV$ decays  have been calculated very recently with the inclusion
of all currently known next-to-leading order (NLO) contributions \cite{xiao18a,xiao18b}. For $B_s \to PP$ decays, the situation is a little  complicated:
\begin{enumerate}
\item[(1)]
In 2004, the $B_s \to \pi\pi$ decay was firstly studied by using the PQCD approach at the leading order (LO) ~\cite{bspipi}.
In 2007,  all thirteen $B_s \to PP $ decays were studied in the PQCD approach at leading order in Ref.~\cite{ali07}.
The large branching ratio for $B_s \to \pi^+\pi^-$ decay as predicted in Refs.~\cite{bspipi,ali07} are confirmed  several years later by both
CDF \cite{CDFpipi} and LHCb measurements \cite{LHCbpipi,LHCbpik}.

\item[(2)]
In 2008,   all $B_s \to PP $ decays were studied in the PQCD approach in Ref.~\cite{bs08} with the inclusion of the NLO contributions from different sources
known at that time:
(a) the NLO Wilson coefficients $C_i(\mu)$ with other relevant functions at the NLO level~ \cite{buras96};  (b) the NLO vertex and quark-loop corrections
~\cite{npb675,nlo05,joint} and (c)  the NLO contribution from the operator $O_{8g}$~\cite{nlo05,o8g2003}.

\item[(3)]
After 2012,    the NLO twist-2 and twist-3 contributions to $B/B_s \to P$ form factors  are calculated in Refs.~\cite{prd85-074004,cheng14a}.
The $B_s \to (K\pi, KK) $ and $B_s \to (\pi \etap,\etap\etap)$ decays are  studied very soon in Refs.~\cite{xiao14a,xiao14b} with the inclusion
of newly known NLO contributions to the relevant form factors.
Although the pure annihilation decays $B_s \to \pi\pi$  do not  receive  the NLO contributions to $B/B_s \to P $ form factors \cite{xiao2012},
the studies for $B_s \to K\etap$ at the same NLO level  as  that for  $B_s \to (K\pi, KK,\pi\etap,\etap\etap) $ decays
~\cite{xiao14a,xiao14b} are very  interesting and worth of  being done now.

\end{enumerate}
From the above mentioned works \cite{xiao2012,xiao14a,xiao14b,xiao18a,xiao18b}, we get to know that (a) the NLO contributions can interfere
with the LO part constructively or destructively for different decay modes and can therefore result in large variations to the  LO  predictions;
and (b)  the agreement between  the PQCD predictions for the decay rates and CP violating asymmetries and those currently
available experimental measurements can be improved effectively after the inclusion of the NLO contributions.

In this paper ,  we will calculate the $B_s \to K\etap$ decays with the inclusion of all currently known NLO contributions, reexamine
other $B_s \to PP$ decays simultaneously by using the same set of wave functions and  input parameters,  compare our  PQCD predictions
with those obtained based on other different approaches, as well as currently available measured values for five
decay modes,  and finally check  the  effects of the NLO contributions.

This paper is organized as follows. In Sec.~\ref{sec:lo-nlo}, we give a brief review about the PQCD factorization approach
and we calculate analytically the relevant Feynman diagrams and present the various decay amplitudes
for the considered decay modes at the LO and NLO level.
We  show the numerical PQCD predictions for the branching ratios and  CP violating asymmetries of all thirteen $B_s \to PP $ decays
in Sec~\ref{sec:n-d} and make phenomenological analysis.  The summary  will be given in Sec.~\ref{sec:4}.

\begin{table}
\caption{ The measured values of the branching ratios (in units of $10^{-6}$) of the five considered decay modes,
as reported by the Belle\cite{bellepipi,bellekk}, CDF\cite{CDFpipi,CDFpik,CDFkk},
LHCb Collaboration \cite{LHCbpipi,LHCbepep,LHCbpik}, and the world averages as given in Refs.~\cite{hfag2018,pdg2018}.}
\label{tab:brexp}
\begin{tabular*}{16cm}{@{\extracolsep{\fill}}l|lll|ll} \hline\hline
{\rm Mode} $(\bar{B}_s^0\to \bar{f} ) $ &{\rm Belle}\cite{bellepipi,bellekk} &{\rm CDF}\cite{CDFpipi,CDFpik,CDFkk}&{\rm LHCb}\cite{LHCbpipi,LHCbepep,LHCbpik}&{\rm HFLAV}\cite{hfag2018}&{\rm PDG}\cite{pdg2018} \\ \hline
$\bar B_s^0\to \pi^- K^+$&      $<26$&     $5.3\pm0.9\pm0.3    $&$5.6\pm0.6\pm0.3$&$5.3\pm0.5  $&$5.7\pm0.6 $\\
$\bar B_s^0\to K^+ K^-$&      $35^{+10}_{-9}\pm7$&     $25.9\pm2.2\pm1.7    $&$23.7\pm1.6\pm1.5$&$24.8\pm1.7 $&$25.9\pm1.7  $\\
$\bar B_s^0\to K^0 \bar K^0$&      $19.6^{+5.8}_{-5.1}\pm2.0$&     $-$&$-$&$19.6^{+6.2}_{-5.6} $&$20\pm6$\\
$\bar B_s^0\to \etar \etar$&      $-$&     $-    $&$33.1\pm7.0\pm1.2$ &$33.1\pm7.1 $&$33\pm7 $\\
$\bar B_s^0\to \pi^+\pi^-$&      $<12$&     $0.60\pm0.17\pm0.04   $&$0.691\pm0.083\pm0.044$ &$0.67\pm0.08$&$0.70\pm0.08 $\\
 \hline\hline
\end{tabular*}
\end{table}

\begin{table}
\caption{ The measured values of  $A_{\rm CP}(B^0_s\to \pi^+ K^-)$ and  CP-violating asymmetries $C_{KK}$,  $S_{KK}$ and $A^{\Delta \Gamma}_{KK}$
of $B^0_s\to K^+ K^-$  decay, as reported by the CDF\cite{CDFpikcp}and LHCb Collaboration \cite{LHCbpikcp1,LHCbpikcp2,lhcb-18a,LHCbkkcp},
and the world averages as given in Refs.~\cite{hfag2018,pdg2018}.}
\label{tab:acpexp}
\begin{tabular*}{16cm}{@{\extracolsep{\fill}}l|ll|ll} \hline\hline
{\rm Mode}                                  &{\rm CDF}\cite{CDFpikcp}       &{\rm LHCb}\cite{LHCbpikcp1,LHCbpikcp2,lhcb-18a,LHCbkkcp}      &{\rm HFLAV}\cite{hfag2018}&{\rm PDG}\cite{pdg2018} \\  \hline
$A_{\rm CP}( B_s^0\to K^-\pi^+)$         &      $0.22\pm 0.07$   &     $0.27\pm0.08\pm0.02  $ \cite{LHCbpikcp1}                             &$0.213\pm0.017$                      & $0.221\pm0.015 $ \\
 &                                                            &     $0.27\pm0.04\pm0.01  $ \cite{LHCbpikcp2}                                               &                                                          &      \\
 &                                                            &     $0.213\pm0.015\pm0.007 $   \cite{lhcb-18a}                                              &                                                         &     \\ \hline
$C_{KK}( B_s^0\to K^+K^-) $                                      &                                                             &     $0.14\pm0.11\pm0.03    $ \cite{LHCbkkcp}         &                                                          &$0.14\pm 0.11 $ \\
                                                         &                                                             &     $0.20\pm0.06\pm0.02    $  \cite{lhcb-18a}                                                    &                                                          & \\ \hline
$S_{KK}( B_s^0\to K^+K^-) $                                      &                                                             &     $0.30\pm0.12\pm0.04    $ \cite{LHCbkkcp}                                                     &                                                          &$0.30\pm 0.13 $ \\
                                                         &                                                             &     $0.18\pm0.06\pm0.02    $      \cite{lhcb-18a}                                                &                                                          & \\ \hline

$A^{\Delta \Gamma}_{KK}( B_s^0\to K^+K^-) $                                      &                                                             &     $-0.79\pm0.07\pm0.10    $ \cite{lhcb-18a}              &          &  \\
 \hline\hline
\end{tabular*}
\end{table}

\section{ Decay amplitudes at LO and NLO level}\label{sec:lo-nlo}

As usual, we consider the $B_s$ meson at rest and treat it as a heavy-light system.
Using the light-cone coordinates, we define the $B_s^0$ meson with momentum
$P_1$, the emitted meson $M_2$ and the recoiled meson $M_3$ with momentum $P_2$
and $P_3$ respectively. We also use $x_i$ to denote the momentum fraction
of anti-quark in each meson and set the momentum $P_i$ and $k_i$ ( the momentum carried by the
light anti-quark in $B_s$ and $M_{2,3}$ meson) in the following forms:
\beq
P_1 &=& \frac{\mbs}{\sqrt{2}} (1,1,{\bf 0}_{\rm T}), \quad
P_2 = \frac{M_{B_s}}{\sqrt{2}}(1,0,{\bf 0}_{\rm T}), \quad
P_3 = \frac{M_{B_s}}{\sqrt{2}} (0,1,{\bf 0}_{\rm T}),\non
k_1 &=& (x_1 P_1^+,0,{\bf k}_{\rm 1T}), \quad
k_2 = (x_2 P_2^+,0,{\bf k}_{\rm 2T}), \quad
k_3 = (0, x_3 P_3^-,{\bf k}_{\rm 3T}).
\eeq
The integration over $k_{1,2}^-$ and $k_3^+$  will lead conceptually to the
decay amplitudes
\beq
\cala &\sim &  \int\!\! d x_1 d x_2 d x_3 b_1 d b_1 b_2 d b_2 b_3 d b_3 \non
&& \vspace{1cm}\cdot \mathrm{Tr}\left [ C(t) \Phi_{B_s}(x_1,b_1) \Phi_{M_2}(x_2,b_2) \Phi_{M_3}(x_3, b_3) H(x_i,
b_i, t) S_t(x_i)\, e^{-S(t)} \right ], \quad \label{eq:a2}
\eeq
where $b_i$ is the conjugate space coordinate of $k_{\rm iT}$, $C(t)$  denotes  the Wilson coefficients evaluated at the scale $t$,
and $\Phi_{B_s}$ and $\Phi_{M_i}$ are wave functions of the $B_s$ meson and
the final state mesons. The hard kernel $H(x_i,b_i,t)$ describes the four-quark
operator and the spectator quark connected by a hard gluon. The Sudakov factors $e^{-S(t)}$ and $S_t(x_i)$ together suppress the soft
dynamics effectively \cite{li2003}.

\subsection{ Wave functions}\label{sec:wf}

Without the endpoint singularities in the evaluations, the hadron wave functions are the only input in the PQCD approach.
These nonperturbative quantities are process independent and could be obtained with the techniques of QCD sum rule and/or Lattice QCD,
or be fitted to the measurements for some relevant decay processes with good precision.

For $B_s^0$ meson, we consider only the contribution of Lorentz structure~\cite{ali07}
\beq
\Phi_{B_s}= \frac{1}{\sqrt{6}} (\psl_{B_s} +m_{B_s}) \gamma_5 \phi_{B_s} ({\bf k_1}),
\label{eq:bsmeson}
\eeq
and adopt the distribution amplitude $\phi_{B_s}$ as in Refs.~\cite{bspipi,ali07,xiao14a}.
\beq
\phi_{B_s}(x,b)&=& N_{B_s} x^2(1-x)^2 \exp \left  [ -\frac{M_{B_s}^2\ x^2}{2 \omega_{B_s}^2}
-\frac{1}{2} (\omega_{B_s} b)^2\right].
\label{phib}
\eeq
We also take $\omega_{B_s} =0.50 \pm 0.05$ GeV in numerical calculations.
The normalization factor $N_{B_s}$ will be determined  through the
normalization condition: $\int_0^1 dx \; \phi_{B_s}(x,b=0)=f_{B_s}/(2\sqrt{6})$.

For $\eta$-$\etar$ mixing, we also use the quark-flavor basis:
$\eta_q= (u\bar u +d\bar d)/\sqrt{2}$ and $\eta_s=s\bar{s}$ \cite{fks98,fks99,xiao08b,fan2013}.
The physical $\eta$ and $\etar$ can then be written in the form of
\beq
\left(\begin{array}{c} \eta \\ \eta^{\prime}\end{array} \right)= \left ( \begin{array}{cc}
\cos\phi & -\sin\phi\\ \sin\phi & \cos\phi\\ \end{array} \right)
\left(\begin{array}{c} \eta_q \\ \eta_s\end{array} \right),\label{eq:e-ep2}
\eeq
where $\phi$ is the mixing angle. The relation between the decay constants
$(f_\eta^q, f_\eta^s,f_{\etar}^q,f_{\etar}^s)$ and $(f_q,f_s)$ can be found for example
in Ref.~\cite{xiao08b}. The chiral masses $m_0^{\eta_q}$ and $m_0^{\eta_s}$ have been defined
in Ref.~\cite{ckl06} by assuming the exact isospin symmetry $m_q=m_u=m_d$.
The three input parameters $f_q, f_s,$ and $\phi$ in Eq.~(\ref{eq:e-ep2}) have been extracted from the data \cite{fks98,fks99}
\beq
f_q=(1.07\pm 0.02)f_{\pi},\quad f_s=(1.34\pm 0.06)f_{\pi},\quad \phi=39.3^\circ \pm 1.0^\circ.
\eeq
With $f_\pi=0.13$ GeV, the chiral masses $m_0^{\eta_q}$ and $m_0^{\eta_s}$ consequently
take the values of $m_0^{\eta_q}=1.07$ GeV and $m_0^{\eta_s}=1.92$ GeV \cite{ckl06}.

For the final state pseudo-scalar mesons $M=(\pi, K,\eta_q,\eta_s)$, their wave functions are the same
ones as those in Refs.~\cite{pball-90,pball-98,pball-99,BL-04,KMM-04,pball-05,pball-06,csbwf1,csbwf2}:
\beq
\Phi_{M_i}(P_i,x_i)\equiv \frac{1}{\sqrt{6}}\gamma_5 \left [ \psl_i \phi^{A}_{M_i}(x_i)
+m_{0i} \phi_{M_i}^{P}(x_i)+ \zeta m_{0i} (\nsl \vsl -1)\phi_{M_i}^{T}(x_i)\right ],
\label{eq:phip}
\eeq
where $m_{0i}$ is the chiral mass of the meson $M_i$,  $P_i$ and $x_i$ are the momentum and
the fraction of the momentum of $M_i$s. The parameter $\zeta=1$ or $-1$ when the
momentum fraction of the quark (anti-quark) of the meson is set to be $x$.
The distribution amplitudes (DA's) of the pseudo-scalar meson $M$ can be found easily in Refs.~\cite{fan2013,xiao08b,xiao18a}:
\beq
\phi_{M}^A(x) &=&  \frac{3 f_M}{\sqrt{6} } x (1-x)
    \left[1+a_1^{M}C^{3/2}_1(t)+a^{M}_2C^{3/2}_2(t)+ a_4^M C_4^{3/2}(t) \right],\label{eq:piw1}\\
\phi_M^P(x) &=&   \frac{f_M}{2\sqrt{6} }
   \left \{ 1+\left (30\eta_3-\frac{5}{2}\rho^2_{M} \right ) C^{1/2}_2(t)
   -3\left [ \eta_3 \omega_3 + \frac{9}{20}\rho_M^2\left ( 1 + 6a_2^M\right)C_4^{1/2}(t)\right]
   \right \}, \ \
\label{eq:piw2}   \\
\phi_M^T(x) &=&  \frac{f_M(1-2x)}{2\sqrt{6} }
   \left\{ 1+6\left [ 5\eta_3-\frac{1}{2}\eta_3\omega_3-\frac{7}{20}\rho^2_M
   -\frac{3}{5}\rho^2_M a_2^{M} \right ]
   \left (1-10x+10x^2\right )\right \},\quad
   \label{eq:piw3}
\eeq
where $t=2x-1$, $f_M$ and $\rho_M$ are the decay constant and  the mass ratio with the definition of
$\rho_M=(m_\pi/m_0^\pi,m_K/m_0^K$, $m_{qq}/m_0^{\eta_q},m_{ss}/m_0^{\eta_s})$. The parameters $m_{qq}$ and $m_{ss}$ have been
defined in Ref.~\cite{ckl06}:
\beq
m_{qq}^2 &=& m_\eta^2 \cos^2\phi + m_{\etar}^2 \sin^2\phi
- \frac{\sqrt{2}f_s}{f_q} (m_{\etar}^2-m_\eta^2) \cos\phi \sin\phi , \non
m_{ss}^2 &=& m_\eta^2 \sin^2\phi + m_{\etar}^2 \cos^2\phi
- \frac{\sqrt{2}f_q}{f_s} (m_{\etar}^2-m_\eta^2) \cos\phi \sin\phi ,
\eeq
with the assumption of exact isospin symmetry $m_q=m_u=m_d$.
The explicit expressions of those Gegenbauer polynomials $C_1^{3/2}(t)$ and $C_{2,4}^{1/2,3/2}(t)$ can be found
for example in Eq.~(20) of Ref.~\cite{xiao08b}
The Gegenbauer moments $a_i^M$ and other input parameters are similar with those as being used in Refs.~\cite{xiao2012,xiao14a,xiao14b}
\beq
a^{\pi,\eta_q,\eta_s}_1 &=& 0,\quad a^K_1  = 0.06, \quad a^{\pi}_2=0.35\pm 0.15, \quad a_2^K=0.25\pm 0.10, \non
a^{\eta_q,\eta_s}_2&=& 0.115\pm 0.115,  \quad  a^{\pi,K,\eta_q,\eta_s}_4  = -0.015, \quad  \eta_3= 0.015, \quad \omega_3=-3.0,
\label{eq:aim}
\eeq
with the chiral  masses $m_0^\pi=1.4\pm 0.1 $ GeV, $m_0^K=1.9\pm 0.2$ GeV \cite{xiao2012}.

\subsection{ Example of the LO decay amplitudes}\label{sec:lo-aml}

In the SM, for the considered $\bar{B}^0_s \to P P$ decays induced by
the $b \to q$ transition with $q=(d,s)$,
the weak effective Hamiltonian $H_{eff}$ can be written as\cite{buras96}
\beq
\label{eq:heff}
{\cal H}_{eff} &=& \frac{G_{F}}{\sqrt{2}}     \Bigg\{ V_{ub} V_{uq}^{\ast} \Big[
 C_{1}({\mu}) O^{u}_{1}({\mu})  +  C_{2}({\mu}) O^{u}_{2}({\mu})\Big]
  -V_{tb} V_{tq}^{\ast} \Big[{\sum\limits_{i=3}^{10}} C_{i}({\mu}) O_{i}({\mu})
  \Big ] \Bigg\} + \mbox{h.c.}
\eeq
where $G_{F}=1.166 39\times 10^{-5}$ GeV$^{-2}$ is the Fermi constant, and
$V_{ij}$ is the Cabbibo-Kobayashi-Maskawa (CKM) matrix element, $C_i(\mu)$
are the Wilson coefficients and $O_i(\mu)$
are the four-fermion operators.
For convenience, the combinations $a_i$ of the
Wilson coefficients are defined as usual~\cite{ali07}:
\beq
a_{1}&=&C_{2}+C_{1}/3,\quad   a_{2}=C_{1}+C_{2}/3,\non
a_{i}&=& \left \{  \begin{array}{ll}
C_{i}+C_{i+1}/3,&  {\rm for} \quad  i=(3,5,7,9), \\
C_{i}+C_{i-1}/3, & {\rm for} \quad  i=(4,6,8,10).\\ \end{array} \right.
\label{eq:ai}
\eeq

\begin{figure}[tb]
\vspace{-5cm}
\centerline{\epsfxsize=18cm \epsffile{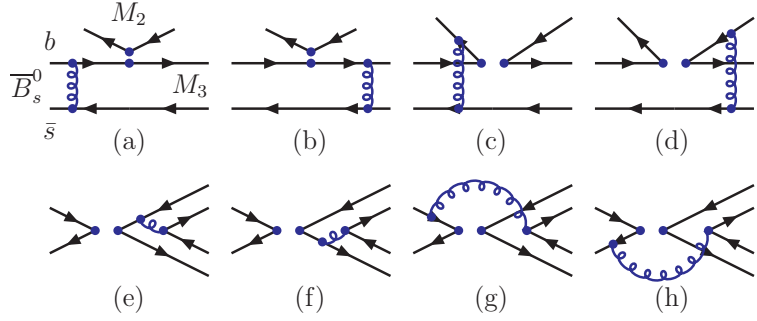}}
\vspace{-16cm}
\caption{ The typical Feynman diagrams which may contribute at leading order to $\bar{B}^0_s \to P P$ decays.}
\label{fig:fig1}
\end{figure}

At leading order, as illustrated in Fig.~\ref{fig:fig1}, there are eight types of Feynman diagrams contributing to the $B_s \to PP$ decays,
which can be classified into three types:
the factorizable emission diagrams ( Fig.~\ref{fig:fig1}(a) and \ref{fig:fig1}(b));
the nonfactorizable emission
diagrams (Fig.~\ref{fig:fig1}(c) and \ref{fig:fig1}(d));
and the annihilation diagrams (Fig.~\ref{fig:fig1}(e)-\ref{fig:fig1}(h)).
As mentioned in the Introduction, the thirteen $B_s \to PP$ modes have been studied at LO or partial NLO in the PQCD approach in
Refs.~\cite{bspipi,ali07,bs08,xiao14a,xiao14b}.
The factorization formulas of  the LO decay amplitudes with various topologies have been presented explicitly for example in Ref.~\cite{ali07}.
Therefore, after the confirmation by our independent recalculations, we shall not collect those analytic expressions here for simplicity.
In this work, we  try  to examine the effects of all currently known NLO contributions to
all thirteen $B_s \to PP$ decay modes in the PQCD approach by using the same set of the input parameters, and compare the PQCD predictions with
those measured values becoming known recently.

\subsection{ The NLO contributions}

\begin{figure}[tb]
\vspace{-5cm}
\centerline{\epsfxsize=18 cm \epsffile{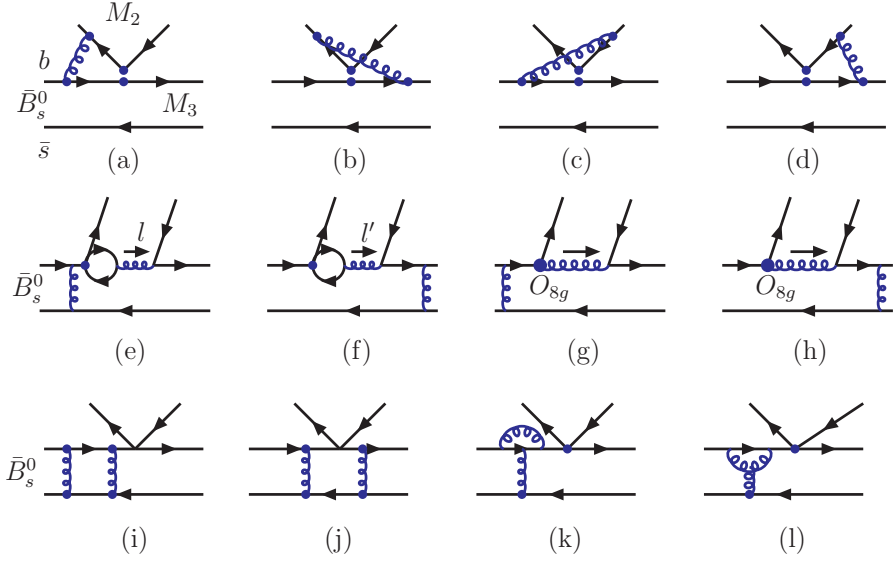}}
\vspace{-13cm}
\caption{Typical Feynman diagrams for NLO contributions:  the vertex corrections (a-d);  the quark-loops (e-f),  the chromo-magnetic penguin
contributions (g-h), and the NLO twist-2 and twist-3 contributions to $B_s \to (K,\eta_s)$ transition form factors (i-l).}
\label{fig:fig2}
\end{figure}

During the past two decades, many authors have made great efforts to calculate the NLO contributions to
the two-body charmless $B/B_s\to M_2 M_3$ in the framework of the PQCD factorization approach.
At present, almost all such NLO contributions become available now:
\begin{itemize}
\item[(1)]
The NLO Wilson coefficients $C_i(m_W)$(NLO-WC),
the renormalization group running matrix $U(m_1,m_2,\alpha)$ at NLO level
and the strong coupling constant $\alpha_s(\mu)$ at two-loop level
as presented in Ref.~\cite{buras96};

\item[(2)]
The NLO contributions from the vertex corrections(VC)~\cite{nlo05,npb675},   the quark-loops(QL)~\cite{nlo05}
and the chromo-magnetic penguin (MP) operator $O_{8g}$ \cite{o8g2003,nlo05},   as illustrated
in Figs.~\ref{fig:fig2}(a)-\ref{fig:fig2}(h).

\item[(3)]
The NLO corrections to the $B_s \to P$ transition form factors,  as shown in Fig.~\ref{fig:fig2}(i)-\ref{fig:fig2}(l).
\end{itemize}
In two previous works \cite{prd85-074004,cheng14a}, we calculated  the  NLO twist-2 and twist-3 contributions to the form factors of $B \to \pi$
transitions.  Based on the $SU(3)$ flavor symmetry, we could extend directly the formulas for the NLO contributions to the form factor
$F_{0,1}^{B \to \pi}(0)$ to the cases for $B_s \to (K, \eta_s) $ transitions after making some proper modifications for the relevant masses
or decay constants of the mesons involved,  as being done in Ref.~\cite{xiao18a} for the decays of $B_s \to PV$.

In this paper, we adopt directly the formulas for all currently known
NLO contributions from Refs.~\cite{npb675,nlo05,o8g2003,fan2013,xiao14a,xiao14b,prd85-074004,cheng14a}
without further discussions about the details.  For the unknown NLO corrections to the nonfactorizable emission
and annihilation decay amplitudes,  however,  some essential comments should be given qualitatively as follows:
\begin{itemize}
\item[(1)]
For the nonfactorizable emission diagrams as shown in Fig.~\ref{fig:fig1}(c,d), since the hard gluons
are emitted from the upper quark line of Fig.~\ref{fig:fig1}(c) and the upper anti-quark
line of Fig.~\ref{fig:fig1}(d) respectively, the LO contribution from these two figures will be
largely cancelled each other. The remaining contribution after the cancellation  will become very small in magnitude.
At NLO level,  with the insertion of second gluon propagator between two quark lines,  another suppression factor $\alpha_s(t)$ will appear.
Of course,  it is worth of mentioning that the  ''Color-suppressed tree" dominated decay modes involving  $\pi^0$ and/or $K^0$ meson,
such as $B_s^0 \to \pi^0 K^0$ decay where  the glauber effects should be considered \cite{Li:2009wba,Li:2014haa,Liu:2015sra,Liu:2016upa},
may be exceptional and need more investigations in depth.
Due to the strong cancellation and the second suppression factor,  in general,  the possible NLO contribution from the spectator diagrams should be much
smaller than the dominant one from the ¡±tree¡± emission diagrams (Fig.~\ref{fig:fig1}(a,b)).

\item[(2)]
For the annihilation diagrams as presented in Fig.~\ref{fig:fig1}(e)-\ref{fig:fig1}(h), the possible NLO contributions
are in fact doubly suppressed by the factors $1/m_{B_s}$ and $\alpha_s(t)$,
and consequently must be much smaller than those dominant LO contribution from Fig.~\ref{fig:fig1}(a) and~\ref{fig:fig1}(b).
\end{itemize}
Therefore, these two kinds of still unknown NLO contributions in the PQCD approach are in fact the higher order corrections to the already small LO pieces, and
should be much smaller than the dominant contribution for the considered decays.

According to Refs.~\cite{npb675,nlo05}, the vertex corrections can be absorbed into the redefinition of the Wilson coefficients
$a_i(\mu)$ by adding a vertex-function $V_i(P)$ to them. The expressions of the vertex-functions $V_i(P)$ can be found easily in Ref.~\cite{nlo05}.
The NLO ``Quark-Loop" and ``Magnetic-Penguin" contributions  are in fact a kind of penguin corrections
with the insertion of the four-quark operators and the chromo-magnetic operator $O_{8g}$ respectively,
as shown in Figs.~2(e,f) and 2(g,h). For the $b\to s$ transition, for example, the corresponding effective Hamiltonian $H_{eff}^{ql}$ and
$H_{eff}^{mp}$ can be written in the following form:
\beq
H_{eff}^{ql}&=&-\sum\limits_{q=u,c,t}\sum\limits_{q{\prime}}\frac{G_F}{\sqrt{2}}
V_{qb}^{*}V_{qs}\frac{\alpha_s(\mu)}{2\pi}C^{q}(\mu,l^2)\left[ \bar{b}\gamma_\rho
\left(1-\gamma_5\right)T^as\right ]\left(\bar{q}^{\prime}\gamma^\rho
T^a q^{\prime}\right),\label{eq:heff-ql}\\
H_{eff}^{mp} &=&-\frac{G_F}{\sqrt{2}} \frac{g_s}{8\pi^2}m_b\;
V_{tb}^*V_{ts}\; C_{8g}^{eff} \; \bar{s}_i \;\sigma^{\mu\nu}\; (1+\gamma_5)\;
 T^a_{ij}\; G^a_{\mu\nu}\;  b_j, \label{eq:heff-o8g}
\eeq
where $l^2$ is  the invariant mass of the gluon which attaches the quark loops
in Figs.~2(e,f), and the functions $C^{q}(\mu,l^2)$ can be found in Ref.~\cite{nlo05,xiao08b}.
The $C_{8g}^{eff}$ in Eq.~(\ref{eq:heff-o8g}) is the effective Wilson coefficient with the definition of $C_{8g}^{eff}= C_{8g} + C_5$ \cite{buras96}.

For the thirteen $B_s\to PP$ decays,  the  analytical evaluations  lead to the following three\Red{\it (two?)} points:
\begin{enumerate}
\item[(1)]
For the $B_s^0 \to \pi^0(\eta, \etar)$ decays,     only the Feynman diagrams Fig.~\ref{fig:fig1}(a)-\ref{fig:fig1}(d) with the $B_s \to \eta_s$ transition will
contribute at leading order.
The relevant NLO contributions are  those from the vertex corrections to the emitted $\pi$ meson and the one  to the $B_s\to \eta_s$ transition
form factor.

\item[(2)]
For the remaining decay modes, besides the LO decay amplitudes, all currently known NLO contributions will contribute in  different ways:
\beq
 {\cal A}^{(u)}_{\pi^- K^+ } & \to &  {\cal A}^{(u)}_{\pi^- K^+  }+{\cal M}^{(u,c)}_{\pi^- K^+ }, \quad
 {\cal A}^{(t)}_{\pi^- K^+  }  \to {\cal A}^{(t)}_{\pi^- K^+  }-{\cal M}^{(t)}_{\pi^- K^+ }-{\cal M}^{(g)}_{\pi^- K^+ },  \non
 {\cal A}^{(u)}_{\pi^0 K^0 } & \to &  {\cal A}^{(u)}_{\pi^0 K^0  }+{\cal M}^{(u,c)}_{\pi^0 K^0  }, \quad
{\cal A}^{(t)}_{\pi^0 K^0   }  \to {\cal A}^{(t)}_{\pi^0 K^0  }-{\cal M}^{(t)}_{\pi^0 K^0 }-{\cal M}^{(g)}_{\pi^0 K^0 }, \non
{\cal A}^{(u)}_{K^+ K^- }  & \to& {\cal A}^{(u)}_{K^+ K^-  }+{\cal M}^{(u,c)}_{K^+ K^-  }, \quad
{\cal A}^{(t)}_{K^+ K^-   }  \to {\cal A}^{(t)}_{K^+ K^-  }-{\cal M}^{(t)}_{K^+ K^- }-{\cal M}^{(g)}_{K^+ K^- },  \non
 {\cal A}^{(u)}_{K^0 \bar K^0 } &\to& {\cal A}^{(u)}_{K^0 \bar K^0 }+{\cal M}^{(u,c)}_{K^0 \bar K^0  }, \quad
{\cal A}^{(t)}_{K^0 \bar K^0   } \, \to\, {\cal A}^{(t)}_{K^0 \bar K^0  }-{\cal M}^{(t)}_{K^0 \bar K^0 }-{\cal M}^{(g)}_{K^0 \bar K^0 }, \non
{\cal A}^{(u)}_{K^0 \eta_{n(s)} } & \to & {\cal A}^{(u)}_{K^0 \eta_{n(s)} }+{\cal M}^{(u,c)}_{K^0 \eta_{n(s)}  }, \quad
 {\cal A}^{(t)}_{K^0 \eta_{n(s)}   } \, \to\, {\cal A}^{(t)}_{K^0 \eta_{n(s)}  }-{\cal M}^{(t)}_{K^0 \eta_{n(s)} }-{\cal M}^{(g)}_{K^0 \eta_{n(s)} }, \non
 {\cal A}^{(u)}_{\eta_s \eta_s }  & \to & {\cal A}^{(u)}_{\eta_s \eta_s }+{\cal M}^{(u,c)}_{\eta_s \eta_s  }, \quad
 {\cal A}^{(t)}_{\eta_s \eta_s   } \to {\cal A}^{(t)}_{\eta_s \eta_s }-{\cal M}^{(t)}_{\eta_s \eta_s }-{\cal M}^{(g)}_{\eta_s \eta_s },
\label{eq:nlo-a}
\eeq
where the terms ${\cal A}_{M_2M_3}^{(u,t)}$ refer to the LO amplitudes,  while ${\cal M}_{M_2M_3}^{(u, c, t )}$
and ${\cal M}_{M_2M_3}^{(g)}$ are the NLO amplitudes, which describe the NLO contributions from the quark-loops,
the QCD-penguin-loops and the magnetic-penguin diagrams, respectively.
The explicit expressions and more details about these NLO amplitudes can be found easily for example in Refs.~\cite{nlo05,xiao14a,xiao14b}.

\end{enumerate}

As mentioned in previous section,  we will extend the formulaes of the NLO contributions for $B\to \pi$
transition form factors as given in Refs.~\cite{prd85-074004,cheng14a} to the cases for $B_s \to (K, \eta_s)$ transition form factors.
The NLO form factor $f^+(q^2)$ for $B_s \to K$ transition, for example, can be written in the following form:
\beq
f^+(q^2)|_{\rm NLO}^{B_s\to K} &=& 8 \pi m^2_{B_s} C_F \int{dx_1 dx_2} \int{b_1 db_1 b_2 db_2}
\phi_{B_s}(x_1,b_1)\non &&
\hspace{-2cm}\times \Biggl \{ r_k \left [\phi_{k}^{P}(x_2) - \phi_{k}^{T}(x_2) \right ]
\cdot \alpha_s(t_1)\cdot e^{-S_{B_s k}(t_1)}\cdot S_t(x_2)\cdot h(x_1,x_2,b_1,b_2) \non
&&\hspace{-2cm}  + \Bigl [ (1 + x_2 \eta)
\left (1 + F^{(1)}_{\rm T2}(x_i,\mu,\mu_f,q^2)\; \right ) \phi_{k}^A(x_2)
+ 2 r_k \left (\frac{1}{\eta} - x_2 \right )\phi_{k}^T(x_2)
- 2x_2 r_k \phi_{k}^P(x_2) \Bigr ] \non
&& \hspace{-1cm} \cdot \alpha_s(t_1)\cdot e^{-S_{B_s k}(t_1)} \cdot S_t(x_2)\cdot h(x_1,x_2,b_1,b_2)\non
&& \hspace{-2cm} + 2 r_{k} \phi_{k}^P(x_2) \left (1 + F^{(1)}_{\rm T3}(x_i,\mu,\mu_f,q^2)\right )
\cdot \alpha_s(t_2)\cdot e^{-S_{B_s k}(t_2)} \cdot S_t(x_2)\cdot h(x_2,x_1,b_2,b_1) \Biggr \},
\label{eq:ffnlop}
\eeq
where $r_k=\Red{\it m_0^K}/m_{B_s}$, $\eta=1-q^2/m_{B_s}^2$ with $q^2=(P_{B_s}-P_3)^2$ and $P_3$ is the momentum of the
meson $M_3$ which absorbed the spectator $\bar s$ quark of the $\bar B_s^0 $ meson, $\mu$ ($\mu_f$) is  the
renormalization (factorization ) scale, the hard scale $t_{1,2}$ are chosen as the largest scale of the propagators in the hard $b$-quark decay diagrams
\cite{prd85-074004,cheng14a}.
The explicit expressions of the threshold Sudakov function  $S_t(x)$ and the hard function $h(x_i,b_j)$ can be found easily in Refs.~\cite{prd85-074004,cheng14a}.
The NLO factors $F^{(1)}_{\rm T2}(x_i,\mu,\mu_f,q^2)$ and $F^{(1)}_{\rm T3}(x_i,\mu,\mu_f,q^2)$ appeared in Eq.~(\ref{eq:ffnlop})
describe the NLO twist-2 and twist-3 contributions to the form factor $f^{+,0}(q^2)$ of the
$B_s \to K$ transition respectively,  and can be written in the following form \cite{prd85-074004,cheng14a}:
\beq
F^{(1)}_{\rm T2}&=& \frac{\alpha_s(\mu_f) C_F}{4 \pi}
\Biggl [\frac{21}{4} \ln{\frac{\mu^2}{m^2_{B_s}}}
-(\frac{13}{2} + \ln{r_1}) \ln{\frac{\mu^2_f}{m^2_{B_s}}}
+\frac{7}{16} \ln^2{(x_1 x_2)}+ \frac{1}{8} \ln^2{x_1} \non
&&+ \frac{1}{4} \ln{x_1} \ln{x_2}
+ \left (- \frac{1}{4}+ 2 \ln{r_1} + \frac{7}{8} \ln{\eta} \right ) \ln{x_1}
+ \left (- \frac{3}{2} + \frac{7}{8} \ln{\eta} \right) \ln{x_2} \non
&&+ \frac{15}{4} \ln{\eta} - \frac{7}{16} \ln^{2}{\eta}
+ \frac{3}{2} \ln^2{r_1} - \ln{r_1}
+ \frac{101 \pi^2}{48} + \frac{219}{16} \Biggr ],  \label{eq:ffnlot2}\\
F^{(1)}_{\rm T3}&=&\frac{\alpha_s(\mu_f) C_F}{4 \pi}
\Biggl [\frac{21}{4} \ln{\frac{\mu^2}{m^2_{B_s}}}
- \frac{1}{2}(6 + \ln{r_1}) \ln{\frac{\mu^2_f}{m^2_{B_s}}}
+ \frac{7}{16} \ln^2{x_1} - \frac{3}{8} \ln^2{x_2} \non
&& \hspace{-1cm}+ \frac{9}{8} \ln{x_1} \ln{x_2}
+ \left (- \frac{29}{8}+ \ln{r_1} + \frac{15}{8} \ln{\eta} \right ) \ln{x_1}
+ \left (- \frac{25}{16} + \ln{r_2} + \frac{9}{8} \ln{\eta} \right) \ln{x_2} \non
&&\hspace{-1cm}+ \frac{1}{2} \ln{r_1} - \frac{1}{4} \ln^{2}{r_1} + \ln{r_2}
- \frac{9}{8} \ln{\eta} - \frac{1}{8} \ln^{2}{\eta} + \frac{37 \pi^2}{32}
+ \frac{91}{32} \Biggr ],
\eeq
where $r_{1,2}=m^2_{B_s}/\xi_{1,2}^2$ with the choice of $\xi_1=25 m_{B_s}$ and $\xi_2=m_{B_s}$. For $B_s \to P P$ decays considered in this paper,
the large recoil region corresponds to the energy fraction  $\eta  \sim \textit{O}(1)$. The factorization scale $\mu_f$ is set to be the hard scales
\beq
t^{a}=\max(\sqrt{x_3 \eta } \, m_{B_s} ,1/b_1,1/b_3), \quad {\rm or} \quad
t^{b}=\max(\sqrt{x_1 \eta } \, m_{B_s} ,1/b_1,1/b_3),\label{tab}
\eeq
corresponding to the largest energy scales in Figs.~\ref{fig:fig1}(a) and \ref{fig:fig1}(b), respectively.
The renormalization scale $\mu$ is defined as \cite{prd85-074004,cheng14a,fan2013}
\begin{eqnarray}
\mu = t_s (\mu_{\rm f})  = \left \{{\rm Exp} \left[ c_1 + \left(\ln
{m_{B_s}^2 \over \xi_1^2}  +{5 \over 4} \right)  \ln{\mu_{\rm f}^2
\over m_{B_s}^2 } \right ]  \, x_1^{c_2 } \, x_3^{c_3} \right \}^{2/21}
\, \mu_{\rm f}, \label{ts function}
\end{eqnarray}
with the coefficients
\begin{eqnarray}
c_1 &=& - \left ({15 \over 4} -{7 \over 16} \ln \eta \right ) \ln
\eta + {1 \over 2} \ln { m_B^2 \over\xi_1^2 }   \left ( 3 \ln { m_B^2 \over\xi_1^2 } + 2 \right ) - {101 \over 48} \pi^2 - {219 \over 16} \,,  \non
c_2 &=& - \left ( 2 \ln { m_B^2 \over\xi_1^2 } + {7 \over 8} \ln \eta - {1 \over 4} \right )  \,, \non
c_3 &=& -{7 \over 8} \ln \eta + {3 \over 2}.
\end{eqnarray}

\section{Numerical results}\label{sec:n-d}

In the numerical calculations, the following input parameters will be used implicitly. The
masses, decay constants and QCD scales are in units of GeV \cite{pdg2018}:
\beq
\Lambda_{\overline{\mathrm{MS}}}^{(f=5)} &=& 0.225,\quad \tau_{B_s^0}=1.51 {\rm ps}, \quad m_b=4.8, \quad M_W = 80.42
\quad  f_{B_{s}} = 0.23\pm 0.02, \quad\non
m_{B_{s}} &=&  5.37,\quad m_K=0.494,\quad  f_K = 0.16, \quad   f_{\pi} = 0.13. \label{eq:para}
\eeq
For the CKM matrix elements, we adopt the Wolfenstein parametrization up to
${\cal O}(\lambda^5)$ with the updated parameters as~\cite{pdg2018}
\beq
\lambda=0.2254\pm 0.0006, \quad A=0.814^{+0.023}_{-0.024}, \quad\bar \rho=0.117 \pm 0.021, \quad\bar \eta=0.353\pm 0.013.
\label{eq:ckm}
\eeq

For the thirteen $B_s^0 \to PP$ decays, their CP-averaged branching ratios are defined  as
\begin{eqnarray}
{ \cal B}&=&\frac{G_F^2\tau_{B_s}}{32\pi  m_{B_s}} \frac{1}{2}\left [ \mid\mathcal{A}(\bar{B}_s^0\to \bar{ f} )\mid^2+
\mid\mathcal{A}(B_s^0\to  f )\mid^2\right ],
\end{eqnarray}
where $\tau_{B_s}$ is the lifetime of the $B_s$ meson.

Since the final state $\bar{f} = K^+ \pi^-$ is not a  CP eigenstate,   the CP asymmetry $\cala_{CP}$ for $\bar{B}_s^0 \to K^+ \pi^-$ decay is defined as
\cite{lhcb-18a}
\beq
A_{\rm CP}(B_s^0 \to \pi^+K^-)= \frac{|\bar{A}_{\bar{f}}|^2 - |A_f|^2}{ |\bar{A}_{\bar{f}}|^2 + |A_f|^2 },
\label{eq:acp-1a}
\eeq
where $A_f$ ($\bar{A}_{\bar{f}}$) is the decay amplitude of $B_s^0 \to f $ ( $\bar{B}_s^0 \to \bar{f} $)  decay.

When the final states are CP eigenstates,  i.e.  $\bar{f} =\eta_f \; f$ with  $\eta_f=\pm 1$ for a CP-even or a CP-odd final state $f$,
the direct CP violation $\cala_f$,  the CP-violating asymmetry ${\cal S}_f$ and $\calh_f$
can be defined in the same way as in Refs.~\cite{ali07,xiao18a,xiao18b}:
\beq
\cala_f=\frac{|\lambda|^2-1}{1+|\lambda|^2}, \quad {\cal S}_f=\frac{2 {\rm Im}[\lambda]}{1+|\lambda|^2},\quad
\calh_f=\frac{2 {\rm Re}[\lambda]}{1+|\lambda|^2}, \label{eq:acpdef1}
\eeq
with the CP-violating parameter $\lambda$
\beq
\lambda=\eta_f  e^{2i\beta_s} \frac{ |A(\bar{B}_s^0 \to f)|}{|A(B_s^0 \to f)|},
\label{eq:lambda}
\eeq
where  $\beta_s =\arg[-V_{ts}V_{tb}^*]$ is small in size for $B_s^0$ meson, and the three  CP violations satisfy the normalization
relation $|\cala_f|^2+|{\cal S}_f|^2+|\calh_f|^2=1$.  It is worth of mentioning that   the parameter
$\cala_f$ and $\calh_f$ as defined in Eq.~(\ref{eq:acpdef1})   have an opposite sign
with  ${\cal C}_f$ and $A_f^{\Delta \Gamma}$ as given in Ref.~\cite{lhcb-18a}:   i.e:  $\cala_f=-{\cal C}_f$ and $\calh_f=- A_f^{\Delta \Gamma}$  .

\subsection{The branching ratios}\label{sec:bsbr}

In Table \ref{tab:br}, we present our numerical results for the CP-averaged branching ratios of the thirteen $\bar B_s^0\to P P$ decays.
In the second column of  Table \ref{tab:br},  we classify   the LO dominant contribution to each decay mode with the symbol  ``${\rm T}$"
(the color-allowed tree), ``${\rm C}$" (the color-suppressed tree), ``${\rm P}$" ( the QCD penguin),
``${\rm P_{EW}}$" ( the electroweak penguin) and ``${\rm ann}$" (the annihilation).
The label ``LO" and ``NLO" denote the PQCD predictions at the leading order only, or with the inclusion of all currently known NLO contributions,
including the NLO twist-2 and twist-3 contributions to the form factors of $B_s \to (K,\eta_s)$ transitions.
The theoretical errors mainly come from the uncertainties of various input parameters, in particular, the dominant
ones come from the shape parameter $\omega_{B_s}=0.50 \pm 0.05$,  the decay constant $f_{B_s}=0.23\pm 0.02$ GeV and the Gegenbauer moments in
the DAs of the relevant mesons. The total errors of the NLO PQCD predictions are given in the Tables
by adding the individual uncertainties in quadrature.
For comparison, we also  show in the fifth to eighth column  of  Table \ref{tab:br}  the LO PQCD predictions as given in Ref.~\cite{ali07},
 the previous PQCD predictions with the inclusion of the partial NLO contributions known at that  time as given in
 Refs. ~\cite{xiao2012,xiao14a,xiao14b},   the  NLO  QCDF predictions as given in Ref.~\cite{npb675} and
the SCET results as given in Ref.~\cite{scet06}.   In last column, we show the currently available measured values for five decay modes
as presented  in PDG 2018 \cite{pdg2018} ( one can see Table \ref{tab:brexp} for more details ).

\begin{table}
\caption{ The LO and NLO PQCD predictions for the CP-averaged branching ratios  (in units of $10^{-6}$) of the considered $\bar B_s^0\to PP$ decays.
The theoretical predictions as given in Refs.~\cite{ali07,xiao2012,xiao14a,xiao14b,bs08,npb675,chengbs09,scet06},  and the world average of the
measured values as given in PDG 2018 \cite{pdg2018}.   }
\label{tab:br}
\begin{tabular*}{16cm}{@{\extracolsep{\fill}}l|l|ll|ll|ll|l} \hline\hline
{\rm Mode}&{\rm Class} &${\rm LO }$&${\rm NLO}$&${\rm PQCD }$ &${\rm PQCD }$ & {\rm QCDF}  &{\rm SCET} & {\rm PDG2018} \\
  ($\bar B_s^0\to \bar{f}$)      &                          & & & \cite{ali07}    &   \cite{bs08,xiao2012,xiao14a,xiao14b}   & \cite{npb675,chengbs09}& \cite{scet06}&\cite{pdg2018}   \\ \hline
$\bar B_s^0\to \pi^- K^+$&      {\rm T}&     $6.9  $&$5.4^{+2.4}_{-1.5}$&$7.6     $&$5.7^{+2.3}_{-1.9} $ \cite{xiao14a}&$10.2 $&$4.9$ & $5.6\pm 0.6$\\ \hline
$\bar B_s^0\to \pi^0 K^0$&      {\rm C}&     $0.18 $&$0.27^{+0.10}_{-0.08}$ &$0.16 $&$0.28^{+0.11}_{-0.07}$\cite{xiao14a}&$0.49$&$0.76$& $-$\\
$\bar B_s^0\to K^0 \eta$&       {\rm C}&          $0.08   $&$0.14^{+0.06}_{-0.04}$ &$0.11 $&$0.19^{+0.04}_{-0.07}$\cite{bs08}&$0.34$&$0.80$& $-$\\
$\bar B_s^0\to K^0 \etar$&      {\rm C}&          $0.60 $&$1.36^{+0.43}_{-0.28}$ &$0.72 $&$1.87^{+0.45}_{-0.56}$\cite{bs08}&$2.0$&$4.5$& $-$\\ \hline
$\bar B_s^0\to K^+ K^-$&      {\rm P}&      $13.4   $&$18.6^{+6.4}_{-5.3}$  &$13.6   $&$19.7^{+6.6}_{-5.7}$\cite{xiao14a}&$22.7$&$18.2$& $25.4\pm 1.7$\\
$\bar B_s^0\to K^0 \bar K^0$&   {\rm P}&      $14.4  $ &$19.7^{+5.9}_{-3.8}$ &$15.6  $&$20.2^{+7.3}_{-5.8}$\cite{xiao14a}&$24.7$&$17.7$ &$20\pm 6$\\
$\bar B_s^0\to \eta \eta$&      {\rm P}&          $6.7 $& $10.4^{+4.9}_{-3.4}$  &$8.0  $&$10.6^{+3.8}_{-2.7}$\cite{xiao14b}&$15.6$&$7.1$& $-$\\
$\bar B_s^0\to \eta \etar$&      {\rm P}&          $17.2   $&$36.2^{+14.7}_{-11.1}$  &$21 $&$41.4^{+16.4}_{-12.0}$\cite{xiao14b}&$54.0$&$24.0$ &$-$\\
$\bar B_s^0\to \etar \etar$&     {\rm P}&        $12.3   $&$30.8^{+11.2}_{-8.6}$     &$14 $&$41.0^{+17.5}_{-13.4}$\cite{xiao14b}&$41.7$&$44.3$& $33\pm 7$\\  \hline
$\bar B_s^0\to \pi^0 \eta$&      ${\rm P_{EW}}$&     $0.04   $&$0.04\pm 0.02$ &$0.05 $&$0.06\pm 0.03$\cite{xiao14b}&$0.075$&$0.014$& $-$\\
$\bar B_s^0\to \pi^0 \etar$&       ${\rm P_{EW}}$&   $0.06   $&$0.07\pm 0.03$ &$0.11 $&$0.13\pm 0.06$\cite{xiao14b}&$0.11$&$0.006$&$-$\\
\hline
$\bar B_s^0\to \pi^+\pi^-$&     {\rm ann}&        $0.62   $&$0.52^{+0.21}_{-0.18}$ &$0.57 $&$0.57^{+0.24}_{-0.22}$\cite{bs08} &$0.02$& $-$ & $0.68\pm 0.08$ \\
                                                       &                         &                        & &&$0.51^{+0.23}_{-0.19}$\cite{xiao2012} & $0.26$\cite{chengbs09}& &\\
$\bar B_s^0\to \pi^0\pi^0$&      {\rm ann}&        $0.25   $ &$0.21^{+0.10}_{-0.09}$  &$0.28$&$0.29\pm 0.12 $ \cite{bs08}&$0.01$&$-$&$-$\\
 \hline\hline
\end{tabular*}
\end{table}

From the theoretical  predictions for the branching ratios of the considered thirteen $B_s \to PP$ decays and those currently available
experimental measurements for the five $B_s \to PP$ decay modes,   as listed in Table \ref{tab:brexp} and   \ref{tab:br},  we have the following observations:
\begin{itemize}
\item[(1)]
For all considered decay channels, the previous LO PQCD predictions of the branching ratios as given in Ref.~\cite{ali07} are well confirmed by our independent
calculations within the errors. The small differences between the LO PQCD  predictions as given in Ref.~\cite{ali07} and in  Table \ref{tab:br}
are induced by the update of  some input parameters.
For most  considered decay channels,  our new NLO PQCD predictions  as listed in the fourth column of Table  \ref{tab:br}  also agree well with those as  given in Refs.~\cite{bs08,xiao2012,xiao14a,xiao14b}.
For ${\cal B}(\bar B_s^0 \to \eta^{\prime} \eta^{\prime})$ decay,  the new PQCD prediction
${\cal B}(\bar B_s^0 \to \eta^{\prime} \eta^{\prime})=(30.8^{+11.2}_{-8.6})\times 10^{-6}$ is smaller than the previous one,  but become agree well
with the measured value:  $(33.1\pm7.0) \times 10^{-6}$~\cite{LHCbepep}. The  reason is that we here used   $a^{\eta_q,\eta_s}_2=0.115$ as input
instead of  $a^{\eta_q,\eta_s}_2=0.44$ as being employed in Ref.~\cite{xiao14b}.

\item[(2)]
For the ``tree" dominated decay $\bar B_s^0 \to \pi^- K^+$, the NLO contribution will result in a $\sim 20\%$ reduction
to the LO PQCD prediction for its branching ratio,  and leads to a better agreement with the data.
The  QCDF prediction ${\cal B}(\bar B_s^0 \to \pi^- K^+)=(10.2^{+6.0}_{-5.2})\times 10^{-6}$ as given in Ref.~\cite{npb675}  is
far above the measured value.  In Ref.~\cite{chengbs09},  however,  the authors presented their QCDF  result
${\cal B}(\bar B_s^0 \to \pi^- K^+)=(5.3^{+0.4+0.4}_{-0.8-0.5}) \times 10^{-6}$ by using  a smaller form factor $F^{B_sK}_0(0)=0.24$
instead of the large one $F^{B_sK}_0(0)=0.31$ in Ref.~\cite{npb675},

\item[(3)]
Among the five  ``QCD-Penguin" decays,   three  decay modes  $\bar B_s^0 \to (K^- K^+, K^0 \bar K^0,  \eta^{\prime} \eta^{\prime})$
have been  measured.   The NLO contributions can provide $\sim 40\%$ to $\sim 150\%$
enhancements to the LO PQCD predictions of the branching ratios  and  help us effectively to obtain a much better agreement between the theory
and the data for these three decays.
Of course,  the QCDF and SCET  predictions for the branching ratios of these three decays as listed in Table \ref{tab:br}   are also
consistent with the experimental measurements within the still large errors.

\item[(4)]
For the three ``Colour-suppressed" decays,  $\bar B_s^0 \to ( K^0 \eta, K^0\eta^\prime,  \pi^0 K^0)$,   the theoretical predictions for their branching ratios are at the
level of $10^{-7}-10^{-6}$,   and have not been observed by experiments.
In PQCD factorization approach, the NLO contributions can provide a factor of  two  enhancement to their decay rates.
The difference between different factorization approaches will be  examined by the future LHCb measurements.

\item[(5)]
The two ``Electroweak-Penguin" decays $\bar B_s^0 \to  \pi^0 \eta^{(\prime)}$ are very rare decay modes,  the theoretical predictions for their decay
rates are at the range of $10^{-8}-10^{-7}$,  and hardly be observed in  near future.
In the PQCD approach, the NLO contributions are  coming  from the so-called ``Vertex corrections" only and lead to a small
enhancement no more than  $20\%$.
The substantial cancelations between the contributions arising from the $u \bar u$ and $d \bar d$ components of the $\pi^0$ meson
is one of the major reasons for so small branching fractions of these two decays.

\item[(6)]
For the two pure annihilation decays $\bar B^0_s \to (\pi^+\pi^-,\pi^0 \pi^0)$,    the NLO correction  comes only   from
 the usage of the Wilson coefficients and their renormalization group evolution at the NLO level,
which  results  in  a $16\%$ reduction to the  corresponding  LO PQCD predictions for their branching ratios.
It is easy to see that although the central value of the  PQCD predictions  for $\calb (\bar B^0_s \to \pi^+\pi^-)$ is a little smaller than the measured one, but
it still  agree well with the measured values within errors.
Although we believe that the still unknown NLO contributions from the annihilation Feynman diagrams  is a higher order
corrections to a small LO quantity,  but it  may  help us to cover the remaining difference between the PQCD prediction and the data.
This is the major motivation for us to complete the calculation for those still unknown NLO pieces.
As is well-known, both the QCDF approach and the SCET  can not provide reliable predictions for  these  pure annihilation decay modes.
In Ref.~\cite{chengbs09}, the authors studied  $\bar B^0_s \to \pi\pi $decays  by  including the subleading power corrections to the penguin annihilation topology,
and gave their prediction ${\cal B}(\bar B^0_s \to \pi^+\pi^-)=(0.26\pm 0.10)\times 10^{-6}$,  which  is  much larger than the one given
in Ref.~\cite{npb675},  but  it is still much smaller  than the measured value.
\end{itemize}

Since the theoretical and experimental errors of the ratios of the branching ratios are generally much smaller than those for the branching ratios themselves,
people tend to define and measure such kinds of ratios.
CDF and LHCb Collaboration , for example,  also defined and measured  some ratios of the branching ratios for several $B/B_s \to PP$ decays
\cite{CDFpipi,CDFpik,CDFkk,LHCbpipi,LHCbpik}  based on some considerations of flavor symmetries ,  as listed in Table  \ref{tab:brexp2}.
\beq
R_1&=&\frac{f_s\calb(B_s^0\to \pi^+\pi^-)}{f_d \calb(B^0\to K^+\pi^-)}, \non
R_2&=&\frac{f_s\calb(B_s^0\to \pi^+\pi^-)}{f_d \calb(B^0\to \pi^+\pi^-)}, \non
R_3&=&\frac{f_s\calb(B_s^0\to \pi^+ K^-)}{ f_d \calb(B^0\to K^+\pi^-)}, \non
R_4&=&  \frac{f_s\calb(B_s^0\to K^+K^-)}{f_d \calb(B^0\to K^+\pi^-)}.
\label{eq:r1234}
\eeq
By employing the PQCD approach, we also calculate the above four ratios  at the LO and NLO level and present our results in the second and third column
of Table  \ref{tab:brexp2}. In the numerical calculations,  $f_s/f_d=0.267^{+0.021}_{-0.020}$ as given in Ref.~\cite{LHCbpik}  is used.
From the PQCD predictions and the measured values as listed in Table \ref{tab:brexp2}, we
find  the following points:
\begin{enumerate}
\item[(1)]  For the ratio $R_1$ and $R_3$, the NLO contributions lead to a significant reduction to the LO results, and such reduction can help us effectively
to explain the measured values. The NLO PQCD results agree well with the corresponding data.

\item[(2)]  For the ratio $R_2$, the NLO contribution is very small in size.  The PQCD predictions for $R_2$ is about half of the measured result, but still
consistent with it within $2\sigma$ error.

\item[(3)]  For the ratio $R_4$, the NLO contribution is also very small in size.  But the PQCD predictions for $R_4$ agree very well
with the measured one within $1\sigma$ error.

\end{enumerate}
It is easy to see that the measured values of $R_{1,2,3,4}$ as listed in Table \ref{tab:brexp2} can be understood properly in the framework
of the PQCD factorization approach at the NLO level.

\begin{table}
\caption{ The LO and NLO PQCD predictions for  some ratios of the branching ratios for  several $B_s/B \to PP $ decays.
The  measured  values of these ratios as reported by CDF \cite{CDFpipi,CDFpik,CDFkk} and LHCb Collaboration \cite{LHCbpipi,LHCbpik},
as well as the averages  from HFLAV \cite{hfag2018}, are listed as comparison. }
\label{tab:brexp2}
\begin{tabular*}{16cm}{@{\extracolsep{\fill}}l|ll|ll|l} \hline\hline
{\rm Mode}& {\rm LO}  & {\rm NLO} & CDF\cite{CDFpipi,CDFpik,CDFkk}& LHCb \cite{LHCbpipi,LHCbpik}& HFLAV \cite{hfag2018} \\ \hline
$R_1 $      & $0.013^{+0.002}_{-0.001} $  & $ 0.008 \pm 0.001$    & $0.008\pm0.002$ & $0.009\pm0.001 $&$0.009\pm0.001 $\\ \hline
$R_2  $      & $0.026 \pm 0.003 $   & $0.027 \pm 0.003$     &$-$ & $0.050^{+0.012}_{-0.011}$&$0.050^{+0.012}_{-0.010}  $\\ \hline
 $ R_3 $         &  $0.149^{+0.013}_{-0.014} $    & $0.087^{+0.011}_{-0.009}$   &  $0.071\pm 0.012$
 & $0.074\pm 0.008 $ & $0.073\pm 0.007 $ \\ \hline
$ R_4 $              &  $ 0.290^{+0.029}_{-0.028}$    & $0.301\pm 0.022 $   &  $0.35\pm 0.03$   & $0.32\pm0.02$   &$0.327 \pm0.017 $\\
 \hline\hline
\end{tabular*}
\end{table}

\subsection{The CP-violating asymmetries}\label{sec:bscpv}

\begin{table}[thb]
\caption{ The LO and NLO PQCD predictions for $A_{CP}(\bar{B}_s^0\to \pi^-K^+)$  and  $\cala_f$  ( in unit of $10^{-2}$ ) for other twelve $\bar B_s^0 \to PP $ decays.
As  a comparison, we also listed  the  theoretical predictions
as given in Refs.~\cite{ali07,bs08,xiao14a,xiao14b,npb675,scet06} and the data as given in Ref.~\cite{pdg2018}.  }
\label{tab:acp1}
\begin{tabular*}{16cm}{@{\extracolsep{\fill}}l|ll|ll|ll|l} \hline\hline
{\rm Mode}& {\rm LO}&{\rm NLO}&${\rm PQCD}$&${\rm PQCD}$&{\rm QCDF}&{\rm SCET}& PDG \\
 $(\bar{B}_s^0\to \bar{f})$       &    & & \cite{ali07}    & \cite{bs08,xiao14a,xiao14b} &\cite{npb675}& \cite{scet06} & \cite{pdg2018} \\ \hline
$\bar B_s^0\to \pi^- K^+$&        $22.9    $&$23.8^{+6.4}_{-6.5}$ &$24.1 $&$38.7 $  \cite{xiao14a}&$-6.7 $&$20$ & $22.1\pm 1.5$ \\ \hline
$\bar B_s^0\to \pi^0 K^0$&       $53.4       $&$87.5^{+6.4}_{-8.3}$  &$59.4$&$83$ \cite{xiao14a}&$41.6$&$76$ & \\
$\bar B_s^0\to K^0 \eta$&           $47.1       $&$95.6^{+3.7}_{-7.6}$  &$56.4$&$96.7$ \cite{bs08}&$46.8$&$-56$ &\\
$\bar B_s^0\to K^0 \etar$&         $-20.6  $&$-42.6^{+2.2}_{-2.0}$ &$-19.9$&$-35.4$ \cite{bs08}&$-36.6$&$-14$ &\\ \hline
$\bar B_s^0\to K^+ K^-$&            $-24.2       $&$-13.2^{+2.3}_{-1.6}$  &$-23.3$&$-16.4$ \cite{xiao14a}&$4.0$&$-6$ & $-14\pm 11$\\
$\bar B_s^0\to K^0 \bar K^0$& $0       $&$0.5\pm 0.1$  &$0$&$-0.7$ \cite{xiao14a}&$0.9 $&$-$&\\
$\bar B_s^0\to \eta \eta$&           $-0.3      $&$-2.7^{+0.4}_{-0.3}$  &$-0.6$&$-2.3$ \cite{xiao14b}&$-1.6$&$7.9$&\\
$\bar B_s^0\to \eta \etar$&         $-0.8      $&$-0.6^{+0.2}_{-0.1}$  &$-1.3$&$-0.2$  \cite{xiao14b}&$0.4$&$0.04$& \\
$\bar B_s^0\to \etar \etar$&       $1.3       $&$2.7^{+0.4}_{-0.3}$  &$1.9$&$2.8$  \cite{xiao14b}&$2.1$&$0.9$&\\ \hline
$\bar B_s^0\to \pi^0 \eta$&        $-4.5         $&$32.2^{+5.3}_{-5.2}$  &$-0.4$&$40.3$  \cite{xiao14b}&$-$&$-$&\\
$\bar B_s^0\to \pi^0 \etar$&     $28.9       $&$59.2^{+2.1}_{-1.5}$  &$20.6$&$51.9$ \cite{xiao14b}&$27.8$&$-$&\\ \hline
$\bar B_s^0\to \pi^+\pi^-$&       $-1.3    $&$-0.6^{+1.0}_{-1.3}$  &$-1.2$&$0.2$ \cite{bs08}&$-$&$-$&\\
$\bar B_s^0\to \pi^0\pi^0$&       $-1.3      $&$-0.6^{+1.0}_{-1.3}$  &$-1.2$&$0.2$ \cite{bs08}&$-$&$-$&\\
 \hline\hline
\end{tabular*}
\end{table}

\begin{table}[thb]
\caption{ The PQCD predictions for the mixing-induced CP asymmetries (in unit of $10^{-2}$)
${\cal S}_f$ (the first row) and $\calh_f$ (the second row) for the twelve $\bar B_s^0 \to PP $ decays.
 In last four columns, we also listed  the  theoretical predictions as given in Refs.~\cite{ali07,xiao2012,xiao14a,xiao14b,chengbs09,scet06}.  }
\label{tab:mixCP}
\begin{tabular*}{15cm}{@{\extracolsep{\fill}}l|ll|ll|ll|l} \hline\hline
{\rm Mode}& {\rm LO }& {\rm NLO} &${\rm PQCD}$&${\rm PQCD}$&{\rm QCDF}&{\rm SCET}& LHCb \\
                      &    &    & \cite{ali07}    &   \cite{bs08,xiao14a,xiao14b}   &\cite{chengbs09}& \cite{scet06} & \cite{lhcb-18a} \\ \hline
$\bar B_s^0\to \pi^0 K^0$&        $-57.1       $&$-45.3^{+8}_{-10}$  &$-61$&$-52.9$  \cite{xiao14a} &$8$&$-16$&\\
                                                 &             $-62.4       $&$-19.2^{+6.3}_{-4.2}$  &$-52$&$-17.4$  \cite{xiao14a} &$-$&$80$& \\
$\bar B_s^0\to K^0 \eta$&     $-52.6       $&$-20.2^{+10.7}_{-15.4}$  &$-43$&$-18$  \cite{bs08} &$26$&$82$ & \\
                                                  &      $-64.2       $&$-20.1^{+12.6}_{-11.5}$  &$-70$&$-18$  \cite{bs08}  &$-$&$7$ & \\
$\bar B_s^0\to K^0 \etar$&   $-65.6  $&$-38.9^{+14.2}_{-13.5}$ &$-68$&$-46$   \cite{bs08}&$8$&$38$ & \\
                                                     &   $-72.1  $&$-81.6^{+7.2}_{-5.4}$ &$-70$&$-82$  \cite{bs08}  &$-$&$-92$ & \\ \hline
$\bar B_s^0\to K^+ K^-$  &             $22.2       $&$23.8^{+5.4}_{-4.2}$  &$28$&$20.6$   \cite{xiao14a} &$22$&$19$ & $18\pm 6\pm 2$\\
                                                 &            $94.4       $&$96.2^{+0.3}_{-1.2}$  &$93$&$96.5$   \cite{xiao14a}  &$-$&$97.9$& $79\pm 7\pm 10$\\
$\bar B_s^0\to K^0 \bar K^0$& $4.0       $&$-3.3$  &$4$&$-0.2$      \cite{xiao14a}     &$0.4$&$-$& \\
                                                        &$\sim 100       $&$\sim 100$  &$\sim 100$&$\sim 100$   \cite{xiao14a}       &$-$&$-$ & \\
$\bar B_s^0\to \eta \eta$&     $3.0  $&$-1.2^{+0.4}_{-0.3}$ &$3$&$-2.2$   \cite{xiao14b}       &$-7$&$-2.6$ &\\
                                                   &      $99.9  $&$99.9$ &$\sim 100$&$99.9$     \cite{xiao14b}       &$-$&$99.6$&\\
$\bar B_s^0\to \eta \etar$&    $3.0  $&$4.1^{+1.0}_{-1.5}$ &$4$&$0.1$   \cite{xiao14b}                &$-1$&$4.1$ &\\
                                                     &    $99.9  $&$99.9$ &$\sim 100$&$\sim 100$    \cite{xiao14b}           &$-$&$99.2$& \\
$\bar B_s^0\to \etar \etar$&  $3.0  $&$2.0^{+0.5}_{-0.3}$ &$4$&$2.5$   \cite{xiao14b}       &$4$&$4.9$ &\\
                                                       &   $99.9  $&$99.9$ &$\sim 100$&$99.9$    \cite{xiao14b}      &$-$&$99.9$ & \\ \hline
$\bar B_s^0\to \pi^0 \eta$&  $11.8      $&$8.7^{+2.6}_{-2.5}$  &$17$&$8$  \cite{xiao14b}   &$26$&$45$ & \\
                                                        &$98.8       $&$94.2^{+1.5} _ {-2.2}$  &$99$&$91.2$  \cite{xiao14b}    &$-$&$-89$ & \\
$\bar B_s^0\to \pi^0 \etar$&$-11.6      $&$-16.9^{+5.2}_{-9.0}$  &$-17$&$-24.9$   \cite{xiao14b}        &$88$&$45$ & \\
                                                        &$95.1       $&$78.8^{+1.0}_{-4.3}$  &$96$&$81.8$ \cite{xiao14b}    &$-$&$-89$& \\ \hline
$\bar B_s^0\to \pi^+\pi^-$&   $11.2 $&$10.6^{+1.2}_{-1.0}$ &$14$&$9$   \cite{bs08}      &$15$&$-$ & \\
                                                       &   $99.9  $&$99.8$ &$99$&$99$    \cite{bs08}        &$-$&$-$ &\\
$\bar B_s^0\to \pi^0\pi^0$&  $11.2  $&$10.6^{+1.2}_{-1.0}$ &$14$&$8.1$  \cite{bs08}         &$15$&$-$ & \\
                                                        &  $99.9  $&$99.9$ &$99$&$99$   \cite{bs08}        &$-$&$-$& \\  \hline\hline
\end{tabular*}
\end{table}

By using the formulaes as given in Eqs.~(\ref{eq:acp-1a},\ref{eq:acpdef1},\ref{eq:lambda}),  we calculate  the direct and mixing-induced CP asymmetries of the thirteen
$\bar B_s^0 \to PP$ decays,  show the numerical results  in Table \ref{tab:acp1}  for $A_{\rm CP} (\bar{B}_s^0\to \pi^-K^+)$ and $\cala_f$ for remaining  twelve decays,
and show  the PQCD predictions for  ${\cal S}_f$ and $\calh_f$  in Table \ref{tab:mixCP} for  twelve $\bar{B}_s^0\to PP$ decays.
As comparison, we also list the theoretical predictions as given in  Refs.~\cite{npb675,scet06} and the data as given in  Refs.~\cite{lhcb-18a,pdg2018}.
From these numerical results we find the following points:
\begin{itemize}
\item[(1)]
Our LO and NLO  PQCD predictions for the direct and mixing-induced CP asymmetries of the
considered $\bar B_s^0 \to PP$ decays  do agree well with those as given in Refs.~\cite{ali07,bs08,xiao14a,xiao14b}.
Some small  differences between the central values are induced by the different choices or upgrade of some input parameters,
such as the Gagenbauer moments and the CKM matrix elements.

\item[(2)]
For most $\bar B_s^0 \to PP$ decays, the  effects of the NLO contributions to the  CP asymmetries
are small  in magnitude.  For $\bar B_s^0 \to  \pi^0 (\etar,K^0)$ and $\bar B_s^0 \to  K^0 \eta^{(\prime)}$ decays,   however,
the NLO enhancements can be as large as $(60-100) \%$.

\item[(3)]
Among the thirteen $\bar B_s^0 \to PP$ decays,  only the CP asymmetries of the $\bar B^0_s \to \pi^- K^+$
and $\bar B^0_s \to K^- K^+$ decays have been measured  by CDF and LHCb Collaboration ~\cite{CDFpikcp,LHCbpikcp1,LHCbpikcp2,lhcb-18a,pdg2018}
as listed in last column of Table   \ref{tab:acp1} and  \ref{tab:mixCP} .
For $\bar B^0_s \to (\pi^- K^+, K^- K^+)$ decays,   fortunately, the NLO PQCD predictions  do agree very well with those
currently  available measured values in both the sign and the magnitude within one standard deviation.

\item[(4)]
For $\bar B^0_s \to (\bar K^0 K^0, \etap\etap,\pi\pi)$ decays,    the CP asymmetries $\cala_f$ and $\cals_f$ are all small in size
and hardly be observed in future experiments.
For  $\bar B^0_s \to  (K^0\pi^0, K^0\etar, \pi^0 \etap)$ decays,  on the other hand,
although their $\cala_f$ and/or  $\cals_f$  may be large in size,   but it is still very  difficult to measure them due to their very small decay rates.

\end{itemize}

\section{SUMMARY}\label{sec:4}

In this paper,  we studied the two-body charmless hadronic decays $\bar B^0_s \to PP$  ( here $P=(\pi, K, \eta, \etar)$)
by employing the PQCD factorization approach with the inclusion of all currently known NLO contributions:
such as the NLO vertex corrections, the quark loop effects, the chromo-magnetic penguin diagrams and
the NLO twist-2 and twist-3 contributions to the relevant form factors $F^{B_s \to K}_0(0)$ and $F^{B_s \to \eta_s}_0(0)$.
In particular, we used  the updated  Gegenbauer moments for the distribution amplitudes of the final state mesons.
We also compared our predictions for the branching ratios and CP violating asymmetries with those currently available experimental
measurements, as well as  the theoretical predictions obtained by using the QCDF approach and SCET method.

By the numerical evaluations and the phenomenological analyses, we found the following interesting points:
\begin{itemize}
\item[(1)]
For  the three $\bar{B}_s^0  \to (K^0 \bar{K}^0, K^+ K^-, \etar \etar)$ decays,  the NLO contributions can provide  about $(40-150) \%$ enhancements to the
LO PQCD predictions for their decay rates.
For ${\cal B}(\bar{B}_s^0  \to \pi^- K^+)$ decay, however,  the NLO contribution will result in a  $ 20\% $ reduction to  the LO PQCD prediction for its branching ratio.
The agreement between the PQCD predictions and the measured values for these three decay modes,  fortunately,  are all improved effectively
after the inclusion of the NLO contributions.

\item[(2)]
For  the pure annihilation $\bar{B}_s^0  \to  \pi^+\pi^-$ decay,  the NLO contribution  will lead to a  $16\%$ reduction to the central value of the LO PQCD prediction.
But the NLO PQCD prediction ${\cal B}(\bar{B}_s^0  \to \pi^+ \pi^-)=(0.52^{+0.21}_{-0.18})\times 10^{-6}$  still agree well with the measured value
$(0.68\pm 0.08) \times 10^{-6}$ \cite{pdg2018} within one standard deviation.

\item[(3)]
Among  the  four ratios of the branching ratios $R_{1,2,3,4}$ defined and measured by CDF  \cite{CDFpipi,CDFpik,CDFkk} and LHCb Collaborations
\cite{LHCbpipi,LHCbpik},  as illustrated in Table \ref{tab:brexp2}, the NLO PQCD predictions for $R_{1,3}$ become agree very well  with the measured
ones after the inclusion of   a  $40\%$ reduction  from the NLO contributions.
The NLO enhancements  to ratio $R_{2,4}$ are very small (less than $4\%$ in size),  the PQCD prediction for $R_4$ agrees very well with the
measured  value,  while the PQCD prediction for $R_2$ is smaller than the measured one but still
consistent with each other within $3\sigma$ errors .

\item[(4)]
For both $\bar{B}_s^0 \to K^+ K^-$ and $\bar{B}_s^0 \to \pi^- K^+$ decays, the NLO PQCD predictions
for the CP-violating asymmetries do agree very well with the measured values \cite{pdg2018} in both the sign and the magnitude.
For the direct CP violation $\cala_f (\bar{B}_s^0\to K^+ K^-)$,  the NLO contribution can  help us to interpret the measured value.

\item[(5)]
For all thirteen $\bar{B}_s^0  \to PP$ decays, we also compared our results with those obtained in the QCDF
and SCET approaches \cite{npb675,sun2003,chengbs09,scet06} and made some comments on the similarities and  the differences
between the theoretical predictions from different approaches.
For most  $\bar{B}_s^0  \to PP$ decays, in fact, the experimental measurements are still absent now.
The forthcoming  precision measurements at LHCb and Belle-II could help us to test  the theoretical predictions.

\end{itemize}

\begin{acknowledgments}
This work is supported by the National Natural Science
Foundation of China under Grants  No.~11775117, 11875033 and No.~11765012 ,
by the Qing Lan Project of Jiangsu Province under Grant No.~9212218405,
and by the Research Fund of Jiangsu Normal University under Grant No.~HB2016004.
\end{acknowledgments}



\end{document}